\documentclass[twocolumn]{aa} % for a paper on 2 column  
\usepackage[varg]{txfonts}
\usepackage{graphicx}
\begin{document}

\title{Increasing the raw contrast of VLT/SPHERE with dark hole techniques\\  III. Broadband reference differential imaging of HR\,4796 using a four-quadrant phase mask\footnote{Based on observations collected at the European Southern Observatory, Chile, 111.24KM}}
\author{Galicher, R.\inst{1}, Potier, A.\inst{2}; Mazoyer, J.\inst{1}; Wahhaj, Z.\inst{3}; Baudoz, P.\inst{1}; Chauvin, G.\inst{4}}
\institute{\inst{1} LESIA, Observatoire de Paris, Université PSL, CNRS, Université Paris Cité, Sorbonne Université, 5 place Jules Janssen, 92195 Meudon, France\\
  \inst{2} Division of Space and Planetary Sciences, University of Bern, Gesellschaftsstrasse 6, 3012 Bern, Switzerland\\
  \inst{3} European Southern Observatory, Alonso de C\'ordova 3107, Vitacura Casilla 19001, Santiago, Chile\\
  \inst{4} Laboratoire Lagrange, Université Côte d’Azur, CNRS, Observatoire de la Côte d’Azur, 06304 Nice, France \\
\email{raphael.galicher at obspm.fr}
}
\authorrunning{R. Galicher et al.}
\titlerunning{SPHERE on-sky Dark Hole with FQPM}
\date\today

% \abstract{}{}{}{}{} 
% 5 {} token are mandatory
  \abstract 
{Imaging exoplanetary systems is essential to characterizing exoplanet atmospheres and orbits as well as circumstellar disks and to studying planet-disk interactions to understand the planet formation processes. Imaging exoplanets or circumstellar disks in the visible and near-infrared is challenging, however, because these objects are very faint relative to their star, even  though only fractions of an arcsecond away. Coronagraphic instruments have already allowed the imaging of a few exoplanets, but their performance is limited by wavefront aberrations. Adaptive optics systems partly compensate for the Earth's atmosphere turbulence, but they cannot fully control the wavefront. Some of the starlight leaks through the coronagraph and forms speckles in the astrophysical image. Focal plane wavefront control, used as a second stage after the adaptive optics system, has been proposed to minimize the speckle intensity within an area called the dark hole.}
{We previously demonstrated the on-sky performance of dark hole techniques, pairwise probing coupled with electric field conjugation, using the apodized pupil Lyot coronagraph of the VLT/SPHERE instrument. In this paper, we probe their performance using the SPHERE four-quadrant phase mask coronagraph, and we demonstrate the interest of combining dark hole techniques and reference differential imaging.}
{We used these dark hole techniques on-sky to create a dark hole in the narrow band around~$1.7\,\mu$m observing HR\,4796. We then recorded broadband images of HR\,4796 and a reference star at the H band.}
{The dark hole techniques improved the H-band detection limit by a factor of three. The dark hole was stable from one star to a nearby star enabling reference differential imaging.}
{This stability offers two new strategies of observation. First, one can quickly create a dark hole observing a bright star before pointing to a faint target star. Furthermore, one can couple dark hole techniques and reference differential imaging. A very interesting point is that the performance of these methods does not depend on the astrophysical signal.}


   \maketitle
%
%________________________________________________________________

   \section{Introduction}
   The detection of exoplanets and circumstellar disks in imaging is challenging because of the faintness and closeness of the objects with respect to their host star. For instance, Earth-like planets can be up to 10 billion times fainter than their host star and separated by fractions of an arcsecond. Coronagraphic instruments, such as VLT/SPHERE \citep{beuzit19}, Gemini/GPI \citep{macintosh14}, Subaru/SCExAO \citep{lozi18}, and Magellan/MagAO \citep{close18}, have been proposed, designed, and built to address this challenge. Despite these advances, only a few young exoplanets have been detected. Indeed, the coronagraph performance is limited by the adaptive optics system's efficiency and the optical aberrations of the instruments. This allows starlight to seep into the instrument and reach the detector, forming speckles in the coronagraphic image \citep{galicher2023}.

      It is worth noting that, even with optimized adaptive optics systems (AO), residual aberrations that remain uncorrected, unseen, or over-corrected will always exist behind ground-based telescopes  \citep{Cavarroc06}. Moreover, both ground-based and space-based coronagraphs will also be limited by wavefront aberrations introduced by optics. Adding a coronagraph by itself behind a telescope will provide images limited by stellar residuals \citep[AO halo and quasi-static speckles][]{galicher2023}. Currently, the speckles are removed using post-processing methods as reference differential imaging \citep[RDI;][]{smith84,beuzit97,lafreniere09, ruane19}. In RDI, one records images of a "reference" star to calibrate the stellar speckles in the target star images. This method, used for SPHERE \citep{Xie2022_rdisphere} and GPI \citep{chen20}, is very useful for extracting images of extended objects that can be severely impacted by other post-processing techniques. Recently, \cite{Wahhaj2021} proposed the star hopping strategy of observation to improve the RDI calibration. The reference star and the target star are observed within minutes, with frequent back and forth to minimize the variation of the speckle pattern. 

      New instruments, such as GPI 2.0 \citep{Chilcote2022} and SPHERE+ \citep{boccaletti2022} for ground-based observatories, or RST/CGI \citep{kasdin2020} in space, are being developed to further reduce the intensity of stellar speckles. They use differential imaging once the raw coronagraphic images are recorded. They also use active optics during the observations to control non-common path aberrations introduced by any optical defects between the astrophysical and the~AO sensing channels. It results in a starlight attenuation in the raw coronagraphic images within an area referred to as the dark hole \citep{galicher2023}. \citet{potier20b} demonstrate that utilizing pair-wise probing \citep{giveon07b} in conjunction with electric field conjugated correction \citep{Malbet1995} results in a ten-fold reduction in starlight intensity between 200\,mas and 700\,mas when using the internal source of the SPHERE instrument. In \citet{potier2022}, we demonstrate the same techniques on-sky to shape the SPHERE deformable mirror (DM) and reduce the starlight between~150\,mas and~600\,mas by up to a factor of~five in the IRDIS images. The most common configuration of SPHERE/IRDIS, which includes an apodized pupil Lyot coronagraph \citep[APLC,][]{Soummer05}, is employed during these demonstrations, and the target star is not known to have any exoplanet or disk. 

      The dark hole techniques used in \citet{potier20b,potier2022} work in close loop. First, one records three to five coronagraphic images introducing several probes, meaning known phase shapes, on the~DM. From this set of images and a model of the light propagation inside the instrument, one calculates the DM shape that minimizes the stellar speckle intensity inside the dark hole area in the coronagraphic image. Then, one starts a new iteration of probing and optimizing the DM shape, and so on. The number of iterations is a trade-off between time used for calibration and time used for astrophysical observation. 
         
   In this work, we used the pair-wise probing and electric-field conjugation techniques, called the dark hole techniques hereafter, with the SPHERE instrument's four-quadrant phase mask coronagraph \citep[FQPM;][]{Rouan00}. First, in Section~\ref{sec:intern_source}, we describe how we enhanced the quality of the FQPM images, which have been hindered by four very bright speckles near the optical axis since the instrument's integration \citep{beuzit19}. We then observed the HR\,4796 system with two main objectives. We aim to demonstrate the usefulness of the dark hole techniques for observing known exoplanetary materials in an on-sky situation. Then, we want to demonstrate that the dark hole techniques can be combined with~RDI methods. We present the data and the method in Section~\ref{sec:obs}. In Section~\ref{sec:DH}, we cover how we measured the gain in contrast that the dark hole techniques offered in the H band on-sky. Furthermore, we demonstrated the dark hole techniques ability to work in conjunction with RDI. We conclude in Section~\ref{sec:conclusion}.

\section{Internal source tests}
\label{sec:intern_source}
Before the on-sky observations, we conducted a daytime investigation of the FQPM coronagraph using the internal source, as we did previously for the APLC~\citep{potier20b}. In the APLC case, we show that the dark hole techniques can remove the diffraction pattern (i.e., structures in the coronagraphic image obtained in the case of no wavefront aberration) and all speckles induced by aberrations. Our aim was mainly to remove the four bright speckles that are visible in the~SPHERE FQPM images at~$\sim2~$resolution elements from the star since the masks were integrated in the instrument~\citep{beuzit19}. In the laboratory, \citet{baudoz18b} demonstrate that~the dark hole techniques can reduce the speckle intensity behind a~FQPM at~two~resolution elements from the optical axis down to~$10^{-5}$ times the non-coronagraphic star intensity using a single~DM. In SPHERE images, we somewhat reduced the intensity of these speckles down to~$\sim5.10^{-3}$ times the non-corongraphic intensity. The reason why we were not able to achieve better performance is not clear. We tried to add strong phase aberrations (first Zernike polynomials up to spherical aberrations) to correct for those speckles. This mainly led to a degradation of the throughput for the off-axis sources. We therefore link the origin of these speckles to defects on the focal plane mask that introduce very strong phase or amplitude aberrations, or to polarization effects; the~SPHERE FQPM images are composed of crossed half-wave plates~\citep{boccaletti08b}. In any case, these four bright speckles will certainly remain in the images and greatly impede performance at very short separations. In the rest of the study, we reduced the speckle intensity further away from the star, between 183\,mas and 625\,mas (see Section~\ref{subsec:rawim}).

\section{Observations}
\label{sec:obs}
In a previous work, we showed the on-sky capabilities of the dark hole techniques, which improved the contrast of raw H3-band ($1667\pm27\,$nm) images by a factor of~five compared to the SPHERE baseline performance \citep{potier2022}. More recently, we carried out two calibration programs to demonstrate the value of the dark hole techniques for on-sky observations of known exoplanetary systems (110.23Z0 and 111.24KM). The primary goal of the new programs was to detect astrophysical signals such as point-like sources (exoplanets or brown dwarfs) or extended sources (circumstellar disks).

During the 110.23Z0 program, we had half a night on October 6, 2022 and one full night on March 31, 2023. The dome remained closed on October 6. The conditions were poor on March 31. It resulted in coronagraphic images that were not speckle limited, whereas this is a requirement of the dark hole techniques. Similarly, no coronagraphic data were obtained during the May, 2023 half-night of the 111.24KM program. Weather conditions were eventually good enough during the~0.4 night of April~7, 2023 (i.e., wind $< 5\,$m$.$s$^{-1}$, no cloud, seeing at~$0.5\,\mu$m $<1"$, and coherence time between $10\,$ms and $15\,$ms). The coronagraphic images were speckle limited, and we were able to probe the dark hole techniques, observing stars with known exoplanets and disks. We observed two well-known systems: Beta Pictoris and HR\,4796. Since the humidity in the instrument rendered the IFS data unusable, we concentrate on IRDIS data in this paper.

As for any observation strategy, one may wonder in which conditions the dark hole techniques are useful. As mentioned above, the techniques need the coronagraph images to be dominated by quasi-static speckles. If the~AO halo is too strong or if the exposure is so short that the AO residuals do not average below the quasi-static speckle level, the dark hole techniques would show a minor impact on the image quality \citep{singh19}. Recently, \cite{Courtney-Barrer2023_EmpiricalContrastModelSPHERE} analyzed and modeled the contrast performance of the SPHERE instrument as a function of the observing conditions. Their model of AO residuals significantly differs from the measured performance when images are not limited by AO. For these cases, they mention a possible limitation to be the quasi-static speckles that they do not take into account. The dark hole techniques that we tested in this work is dedicated to overcoming this speckle limitation. Using their analysis (mainly Figure~7), we conclude that the dark hole techniques are most efficient for bright targets (magnitude in G between 0 and 5) if the weather is better than the median observing conditions and, for mid targets (magnitude in G between 5 and 9, as for HR\,4796), for the best observing conditions (first decile). We note that, even in worse conditions, the dark hole techniques minimize the pinned speckles which the intensity is linked to both the AO residuals and the quasi-static speckles. Hence, GPI2.0 \citep{Chilcote2022} and SPHERE+ \citep{boccaletti2022} will increase the~AO system performance, increasing the amount of time when images are limited by quasi-static speckles. 

During the observation of Beta Pictoris, the airmass was~$1.2$ at sunset, $1.3$ at twilight, and~$1.6$ at the end of the observations. We created a dark hole close to the star to search for planet~c~\citep{lagrange2019}. However, this observation was deemed too challenging due to the presence of the four bright speckles discussed above and a high airmass, which indicates varying atmospheric turbulence. We expended significant effort analyzing the data. However, we obtained no compelling results. Therefore, this paper focuses on the observations of the HR\,4796 system, for which a series of important publications have recently studied the disk with direct imaging in intensity and polarization with GPI and SPHERE \citep{Perrin2015,milli17b,Milli2019,chen20,Olofsson2020,Schmid2021,Xie2022_rdisphere}.

The HR\,4796 system was observed without an apodizer, using the ST$\_$4QPM2 focal plane four-quadrant phase mask and the MASK$\_$STOP$\_$4QPM2 Lyot stop. To stabilize the speckle pattern in the coronagraphic image, we maintained a fixed pupil, enabling the field of view to rotate as in angular differential imaging \citep{marois06}. We first spent~38\,minutes executing five iterations of dark hole correction observing HR\,4796 and following the procedure outlined in~\citet{potier2022}. The correction was carried out using H3 images with~32\,s exposures and five images per iteration, that is, the coronagraphic image and four probing images as explained in~\citet{potier20b,potier2022}. Iteration~0 corresponds to the SPHERE baseline image, which is the image that SPHERE would normally produce using a baseline shape for the~DM. At the final iteration, we recorded the DM shape so that we can apply it at any time. If the system is stable, applying this DM shape will create a dark hole in the coronagraphic image. During the 38\,minutes, we had to redo parts of iterations~2 and~4 (a few images were useless because of bad seeing), which took five minutes. Each iteration took~$\sim8$\,minutes, because we checked multiple parameters to ensure the convergence. Once the process is made available to the community with automatic verification, it can be shortened to $\sim5$\,minutes per iteration. This means that the entire process would take~$\sim25$\,minutes depending on the weather conditions. The longer the lifetime of the~AO residuals, the longer the exposures to average the AO halo~\citep{singh19,potier2022}.

After creating the dark hole at~H3, we recorded 8\,s exposure IRDIS and IFS data at the H band by pointing successively toward~HR\,4796 and a reference star $\sim9'$ away (HD\,109536), using the dark hole~DM shape and the SPHERE baseline DM shape (with no dark hole as SPHERE usually runs). The airmass was~1.1. Table~\ref{tab:seq} presents the starting and ending time of each sequence.
\begin{table}
\centering
\begin{tabular}{cccc}
\multicolumn{4}{c}{Observation sequences}\\
Starting time&End time&Star&DH(y/n)\\
2:43&2:47&HR\,4796&y\\
2:50&2:59&HD\,109536&y\\
3:02&3:17&HR\,4796&y\\
3:23&3:31&HD\,109536&y\\
3:32&3:38&HD\,109536&n\\
3:40&3:51&HR\,4796&n\\
\end{tabular}
\caption{HR\,4796 and HD\,109536 sequences: starting and ending time (UTC). The DH column specifies if the dark hole was applied (y) or not (n).}
\label{tab:seq}
\end{table}
%%
%\begin{table}
%\centering
%\begin{tabular}{ccc}
%\multicolumn{3}{c}{Observation sequences}\\
%Starting time&End time&Star\\
%2:42&2:47&HR\,4796\\
%2:50&2:58&HD\,109536\\
%2:58&3:16&HR\,4796\\
%3:22&3:26&HD\,109536\\
%\end{tabular}
%\caption{\it HR\,4796 and HD\,109536 sequences: starting and end time (UTC) and star.}
%\label{tab:seq}
%\end{table}
%%
For HR\,4796, the total integration time and field-of-view rotation are~$896\,$s and~16$^\circ$.

\section{dark hole on-sky performance at the H band using FQPM}
\label{sec:DH}
In this section, we present raw coronagraphic images recorded using IRDIS/SPHERE at the H band. We quantify the increase in contrast within the dark hole (Section~\ref{subsec:rawim}). Additionally, we establish the on-sky stability of the dark hole, enabling the coupling of a dark hole with~RDI (section~\ref{subsec:rdi}).

\subsection{Raw coronagraphic images}
\label{subsec:rawim}
\begin{figure*}[!ht]
  \centering
  \includegraphics[width=\textwidth]{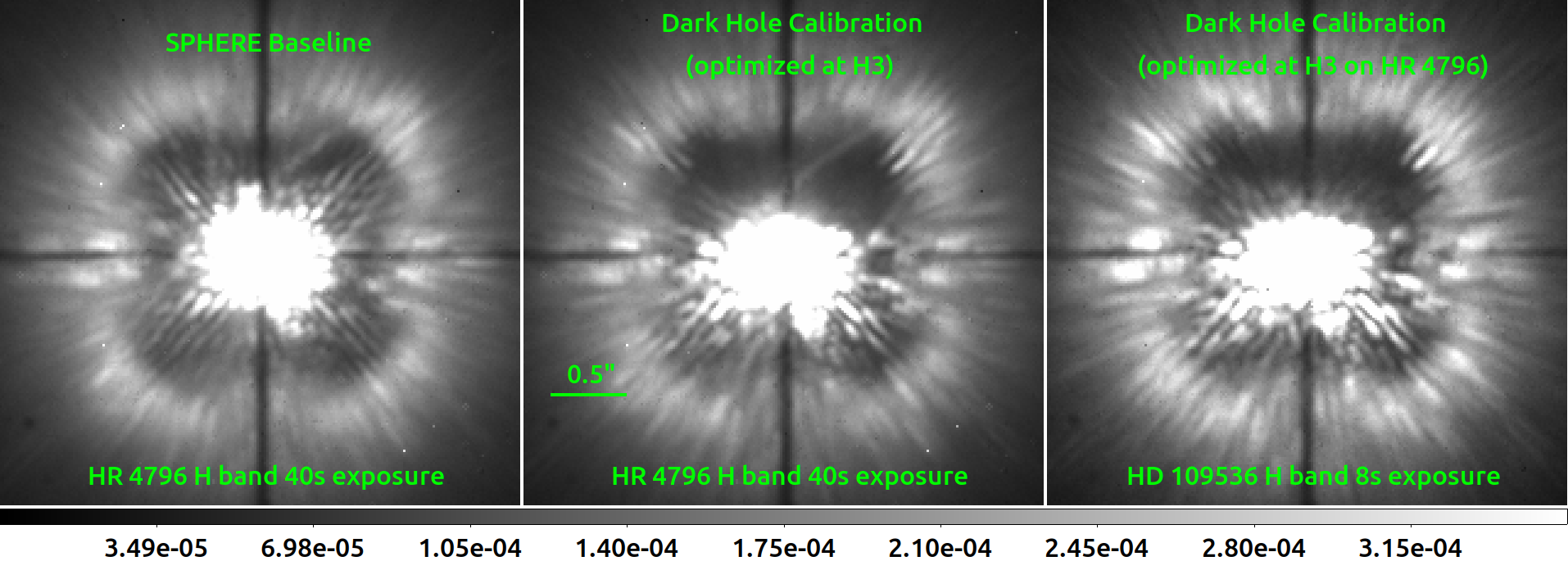}
  \caption{H-band IRDIS 40s images obtained with the SPHERE baseline configuration (left) and after the dark hole calibration optimized at~H3 band (center). The right image is the H-band IRDIS 8s image of the reference star HD\,109536 using the dark hole calibration optimized on HR\,4796. The color bars give the intensity normalized to the maximum of the star image recorded with no coronagraph. The color bar and the spatial scale are the same for the three images.}
  \label{fig:raw_im_h}
\end{figure*}

\begin{figure}[!ht]
  \centering
  \includegraphics[width=.45\textwidth]{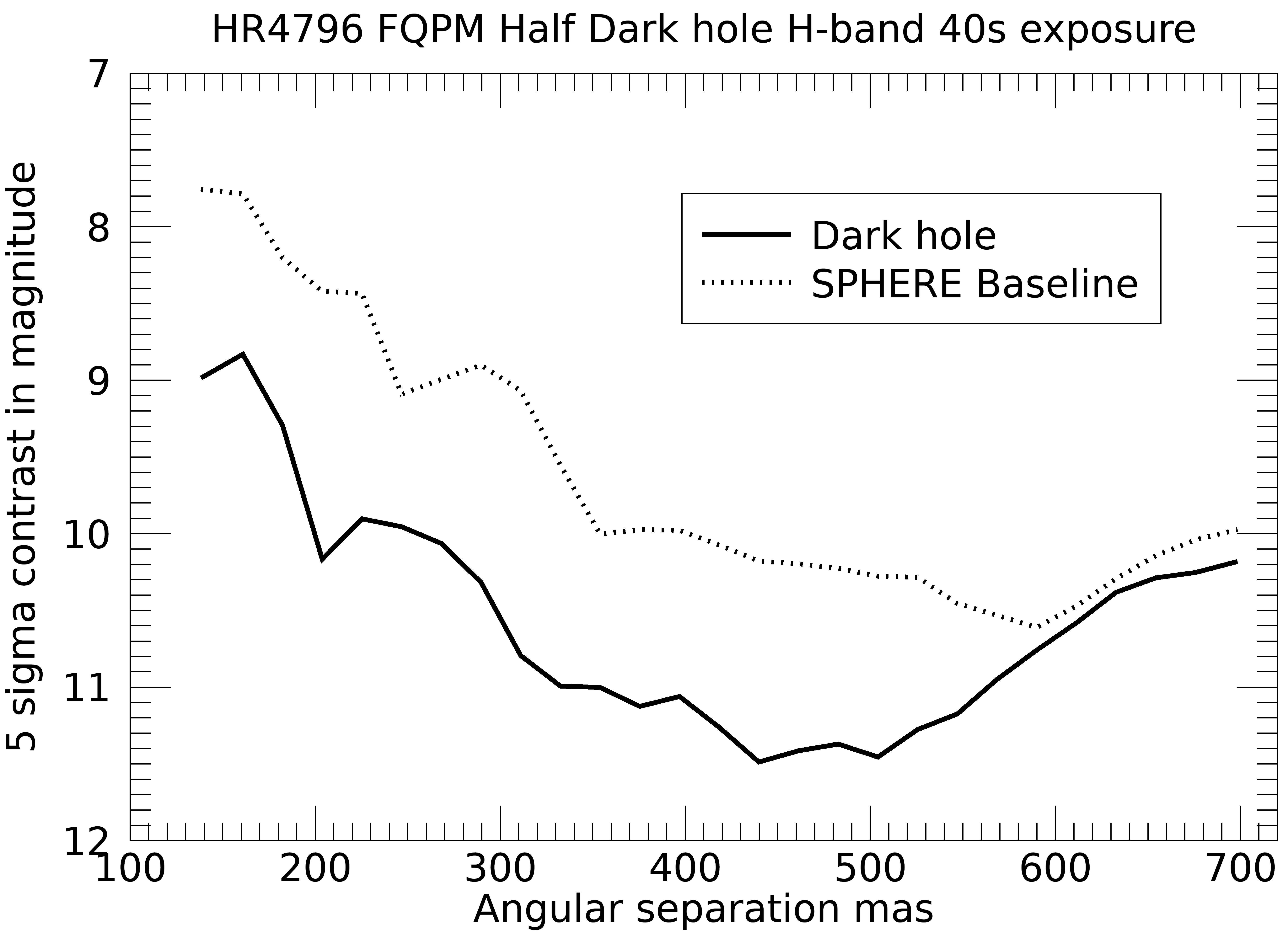}
  \caption{$5\,\sigma$ azimuthal standard deviation as a function of the angular separation from the star for images of Figure~\ref{fig:raw_im_h}: SPHERE baseline (dots) and dark hole calibration (full line).}
  \label{fig:raw_contrast_h}
\end{figure}

Figure~\ref{fig:raw_im_h} displays a single raw coronagraphic image captured by the IRDIS detector with a ~40\,s exposure, using the SPHERE baseline configuration of the~DM (left) or the dark hole DM shape obtained at~H3 (center).
The starlight (speckles and diffraction patterns) is attenuated inside the dark hole. This region spans from~183\,mas to~625\,mas from the star in the upper part of the IRDIS image, removing the 183\,mas-width horizontal area above the star. The circumstellar disk is clearly detected, while it is blended with starlight in the SPHERE baseline image (left). Both images display a black cross, which is a known artifact of the FQPM coronagraph and where no flux is transmitted on the detector. Figure~\ref{fig:raw_contrast_h} plots the contrast curves of the Figure~\ref{fig:raw_im_h} images inside the dark hole area. The contrast curve equals five times the robust standard deviation \citep[that removes deviating values of the distribution;][]{beers1990} of the intensity calculated inside annuli of~$1\,\lambda/D$ width.
The contrast level in the raw on-sky coronagraphic images at the H band is improved by a factor of three inside the dark hole. We also see that starlight remains in the bottom left of the dark hole. We suspect that this light is due to atmospheric phase aberrations that are not well corrected by the AO and that vary too rapidly to be caught by the dark hole loop (wind-driven halo or seeing variation during the exposure).

\subsection{Combining dark hole and reference differential imaging}
\label{subsec:rdi}
Following~\citet{singh19}, we used long exposures to average the~AO residuals and minimize the quasi-static starlight (quasi-static speckles and diffraction pattern) within the dark hole. Consequently, the smooth AO halo becomes the main feature inside the dark hole~(Figure~\ref{fig:raw_im_h}). To eliminate this halo, we can apply an unsharp mask by subtracting the local median from each pixel. Nevertheless, this reduces the circumstellar disk intensity and does not remove the star speckles outside the dark hole. A possible solution to such caveat is to use~RDI.

The effectiveness of combining RDI with dark hole correction needs to be demonstrated. Since the dark hole techniques compensate for optical aberrations caused by the telescope and the instrument, the crucial question is whether the dark hole correction remains stable from the target to the reference star, allowing for stable calibration of the starlight distribution (AO halo, speckles, and diffraction pattern). If the aberrations change, the shape of the~DM that creates the dark hole on the target will not be suitable for the reference star, and new speckles will appear inside the dark hole for this star. This will lead to differences in the images of the reference star and the target, rendering~RDI ineffective.

\begin{figure}[!ht]
  \centering
  \includegraphics[width=.45\textwidth]{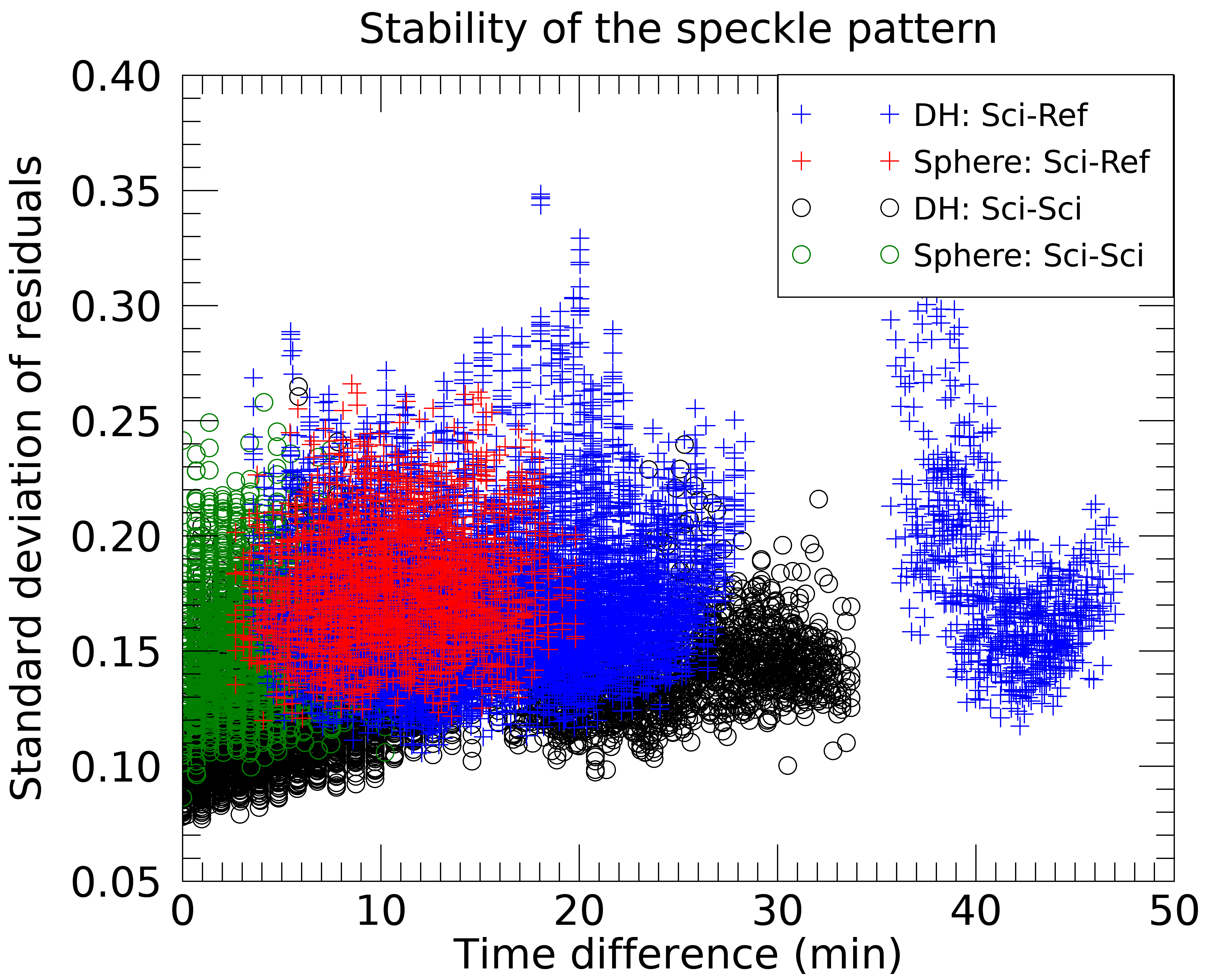}
  \caption{Standard deviation of residuals in difference between individual images as a function of time. Differences are calculated between target frames using the SPHERE baseline images (green) or the dark hole images (black). They are also calculated between target and reference images using the SPHERE baseline images (red) and the dark hole images (blue).}
  \label{fig:stability}
\end{figure}

Sequences of observations for HR\,4796, the target star, and the reference star (HD\,109536, $\sim1^\circ$ away) are recorded to demonstrate the stability of the dark hole correction for use in RDI. One of the 8\,s exposure IRDIS images of HD\,109536 is shown on the right of Figure~\ref{fig:raw_im_h}. The dark hole is well maintained, and the starlight residuals are very similar to the ones observed in the HR\,4796 image (center). We estimate the stability as follows. We call~$I_i$ the~$i^\mathrm{th}$ individual image of HR\,4796 and $I_{R,p}$ the~$p^\mathrm{th}$ individual image of the reference star. For each~$i$ and~$p$ combination, we calculate the standard deviation inside the dark hole region for the difference of the images as follows:
\begin{equation}
v(i,p) = \mathrm{STD}\left(\frac{I_i}{\mathrm{mean}(I_i)}-\frac{I_{R,p}}{\mathrm{mean}(I_{R,p})}\right)
\label{eq:criterium}
.\end{equation}
\noindent The normalization by the average over the dark hole compensates for putative variations of the star flux due to atmospheric conditions. Figure~\ref{fig:stability} plots~$v$ as a function of the time difference between the two images~$I_i$ and~$I_{R,p}$ (blue crosses). The average of~$v$ over all~$i$ and~$p$ values is~$(17.0\pm3.0)\,\%$, meaning that, on average during our sequences, $80.0\,\%$ to~$86.0\,\%$ of the residual intensity inside the dark hole is stable between the target and the reference star. As a comparison, the values of~$v$ calculated in the baseline SPHERE case are plotted in red (no dark hole applied). The average of~$v$ is~$(17.8\pm2.6)\,\%$ in this case, which is slightly higher than for the dark hole case.
The~$v$ values for differences of target images (replacing~$I_{R,p}$ by~$I_p$ in equation~\ref{eq:criterium}) are plotted in black for the dark hole case and in green for the SPHERE baseline case. Again, the residuals are smaller in the dark hole case. Therefore, angular differential imaging would be more efficient using dark hole images than SPHERE baseline images. In this study, the time difference does not span the same range for all configurations, and the number of images is different. Hence, further investigation is needed to properly quantify the stability as a function of time, atmospheric conditions, and angular separation between the target and the reference star. Such a study will be interesting to choose the reference star in an optimal way. However, we find from Figure~\ref{fig:stability} that the dark hole configuration is stable enough to apply~an RDI algorithm.
 
\begin{figure*}[!ht]
  \centering
    \includegraphics[width=\textwidth]{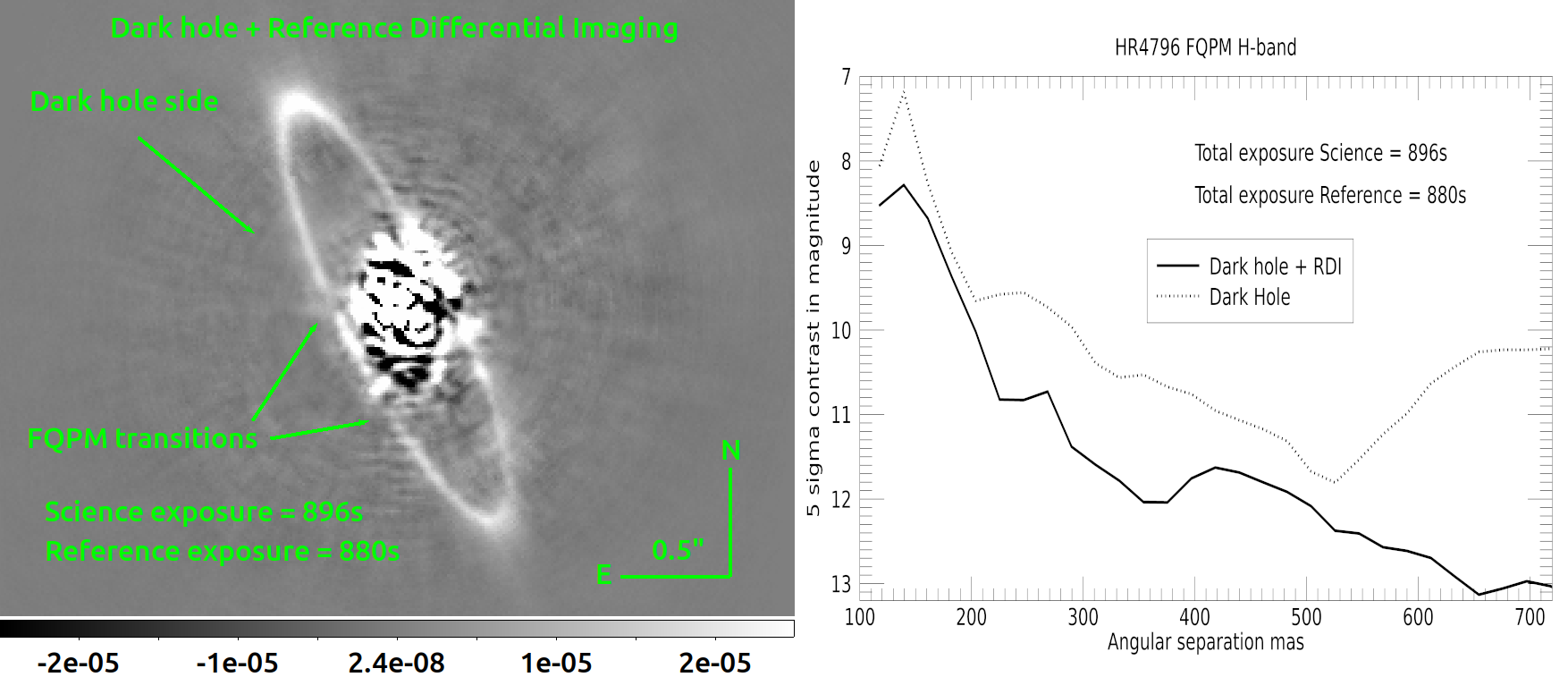}
  \caption{Combination of RDI and dark hole techniques. Left: Resulting image of HR\,4796 circumstellar disk after~RDI using the dark hole calibration on the east side for both the target (896s of integration time) and the reference star (880s of integration time). Two arrows point toward the areas where the intensity is attenuated by the FQPM coronagraph transitions. Right: $5\,\sigma$ azimuthal standard deviation as a function of the angular separation from the star. The standard deviation is calculated in the east part of the image (rough position of the dark hole after rotation of frames). The full line is calculated after~RDI is used, and the dotted line is calculated before.}
  \label{fig:rdi}
\end{figure*}

The two dark hole sequences (HR\,4796 and HD\,109536) are presented in Section~\ref{sec:obs}. One data cube is created for HR\,4796 and another for the reference star. The best linear combination of the reference images that reproduces~$I_i$~($\chi^2$ criterion) is calculated for each image~$I_i$ in the HR\,4796 data cube. The optimization is performed within~$1103\,$mas of the star, removing the pixels within the ellipse of semi-minor and semi-major axes of~368\,mas and~1348\,mas and in the same direction as the~HR\,4796 disk. The ellipse is used to remove any flux from the circumstellar disk that may bias the optimization. It also removes the central part of the image where the starlight intensity can strongly vary from one image to the other because of strong variations of low-order aberrations. The elliptic mask is the same for all images. It is wide enough to hide the disk in all images because the field of view only rotates by~$16^\circ.$  We tried smaller masks (hiding the disk completely) that rotate from one image to the other. The final result is very similar to the one presented below. After the optimization, we subtract each best linear combination from each~$I_i$. We rotate each resulting image so that north is up, and we calculate the median image, which is shown in Figure~\ref{fig:rdi}.
All starlight residuals are well calibrated by~RDI. This means that the dark hole correction that is done on the east side is stable when the telescope is moved from the target star to the reference star; no speckle appears inside the dark hole, and the AO halo inside the dark hole and the speckles and diffraction pattern outside the dark hole are stable. The quality of the image (details of the disk structures) is better on the east side than on the west side because RDI has less to calibrate in the dark hole region; that is, there are fewer speckles inside the dark hole than on the other side (see Figure\,\ref{fig:raw_im_h}). 
The~FQPM transition effects are rotating with the field of view, but they are still noticeable in Figure~\ref{fig:rdi} (left) due to the small rotation of the the field-of-view rotation~($16^\circ$).

The $5\,\sigma$ contrast curve of the residuals on the eastern part of the image before (dotted line) and after (solid line) RDI is plotted on the right of Figure~\ref{fig:rdi}. The robust standard deviation ensures that the estimated standard deviation is not biased by the presence of the disk. Applying~RDI inside the dark hole improves the contrast level by~$\sim1$ magnitude. This demonstrates that~RDI and dark hole techniques can be combined to calibrate the starlight in coronagraphic images: the dark hole techniques actively reduce the residuals during the observations, and~RDI calibrates the residual light a posteriori (mainly the AO halo inside the dark hole). The "one-magnitude improvement" should not be taken as the best calibration RDI can do on the dark hole data, because we only used 22 reference images. Using more reference images and a longer integration time on the target should provide deeper contrast levels. 

Two papers present results of this disk in RDI with SPHERE \citep{Xie2022_rdisphere} and GPI \citep{chen20}. However, they use all the observations of their instrument archive to select the several hundreds most optimized reference images and the comparison to the Figure~\ref{fig:rdi} image is not fair. In the current work, we did not aim to measure the best performance that dark hole and~RDI methods can achieve together. Our objective was to demonstrate that the dark hole techniques can be combined with~RDI. Now this is done, a new strategy of observations can be foreseen to benefit from each technique: the dark hole techniques to minimize the residual starlight in the raw images, the star hopping to optimize the~RDI during one night, and the use of large archives to optimize the RDI calibration over a long survey \citep{chen20,Xie2022_rdisphere}.

\section{Conclusion}
\label{sec:conclusion}
In this study, we first wanted to improve the performance of the SPHERE FQPM coronagraph. Then, we planned to demonstrate the interest of dark hole techniques in observing known exoplanetary systems alone or combined with~RDI.

Using the internal source of the SPHERE instrument, we show that the SPHERE FQPM has defects that cannot be compensated by the~DM of the instrument; the four bright speckles close to the optical axis remain and limit the performance of the SPHERE FQPM, as shown in \citet{beuzit19}. However, we demonstrate that we can improve the performance between $183\,$mas and $625\,$mas.

We also observed the HR\,4796 star with the IRDIS detector using the FQPM coronagraph for half a night. We minimized the starlight residuals inside a dark hole using the pair-wise probing and electric-field conjugation techniques at the H3 band on-sky. The dark hole spanned from~183\,mas to~625\,mas from the star. We demonstrate that this calibration remains accurate enough while expanding the wavelength in the full H-band, and we reduced the starlight by a factor of three. This shows that the dark hole techniques can work on-sky with different kinds of coronagraphs \citep[FQPM in this paper and APLC in][]{potier2022}, and the correction optimized in a narrow band (H3) improves the wide-band image (H band).

Finally, we demonstrate that the dark hole is stable from one star to another, allowing speckle calibration on a bright source before observing the science star (for a fast dark hole calibration), or enabling a combination with~RDI methods such as star hopping~\citep{Wahhaj2021}. In our sequences, we measured that more than~$80\,\%$ of the residual intensity inside the dark hole is stable between the target and the reference star. We also found that the speckle pattern is slightly more stable in the dark hole images than in the SPHERE baseline images (with no dark hole). Further investigation is needed to quantify the stability as a function of critical parameters such as time, atmospheric conditions, or angular separation between the target and the reference star. Combining dark hole and RDI methods may be a new strategy to achieve deep-contrast imaging:  dark hole to actively reduce the residual starlight before the observations and~RDI to calibrate the residual light a posteriori (mainly the AO halo inside the dark hole and speckles and diffraction pattern outside the dark hole). These techniques are very attractive from an astrophysical point of view because they do not modify the astrophysical signal (no self-subtraction of the exoplanet or circumstellar disk image) and they can both probe any angular separation from the star (no limitation as in angular or spectral differential imaging). Furthermore, the performance of these two techniques (RDI and dark hole) does not depend on the astrophysical signal properties, unlike with polarization differential imaging or spectral differential imaging.

\section{Acknowledgment}
This study was supported by the IdEx Université Paris Cité, ANR-18-IDEX-0001. Part of this work was supported by the Physics department of Université Paris Cité. This work was supported by the Action Spécifique Haute Résolution Angulaire (ASHRA) of CNRS/INSU co-funded by CNES. This work was supported by an ECOS-CONICYT grant (\#C20U02).

\bibliographystyle{aa}   %>>>> makes bibtex use spiebib.bst
\bibliography{aa}   %>>>> bibliography data in report.bib

\end{document}